\documentclass{article}

\usepackage{fullpage}
\usepackage{natbib}
\usepackage{authblk}

\usepackage{amsmath,amsfonts,bm}

\def\eqref#1{equation~\ref{#1}}

\def\1{\bm{1}}

\DeclareMathAlphabet{\mathsfit}{\encodingdefault}{\sfdefault}{m}{sl}
\SetMathAlphabet{\mathsfit}{bold}{\encodingdefault}{\sfdefault}{bx}{n}

\usepackage[utf8]{inputenc}
\usepackage[T1]{fontenc}
\usepackage{url}
\usepackage{amsfonts}
\usepackage{multirow} 
\usepackage{multicol}
\usepackage{subcaption}
\usepackage{cleveref}
\usepackage{xurl}
\usepackage{graphicx}
\usepackage{tikz}
\usepackage{blindtext}
\usepackage{caption}
\usepackage{booktabs}
\usepackage{pgf-pie}  

\newcommand{\eg}{\textit{e.g.}}
\newcommand{\ie}{\textit{i.e.}}

\makeatletter

\usepackage{lipsum}

\title{Improving Wikipedia Verifiability with AI}

\author{Fabio Petroni\textsuperscript{1}\thanks{Equal contribution.}\;\,}
\author{Samuel Broscheit\textsuperscript{2}\thanks{Work done during an internship with Meta AI.}\;\,$^*$}
\author{Aleksandra Piktus\textsuperscript{1}}
\author{Patrick Lewis\textsuperscript{1}}
\author{\authorcr Gautier Izacard\textsuperscript{1,4,5}}
\author{Lucas Hosseini\textsuperscript{1}}
\author{Jane Dwivedi-Yu\textsuperscript{1}}
\author{Maria Lomeli\textsuperscript{1}}
\author{Timo Schick\textsuperscript{1}}
\author{Pierre-Emmanuel Mazaré\textsuperscript{1}}
\author{Armand Joulin\textsuperscript{1}}
\author{Edouard Grave\textsuperscript{1}}
\author{Sebastian Riedel\textsuperscript{1,3}}

\affil{\textsuperscript{1}Meta AI, \textsuperscript{2}Amazon Alexa AI, \textsuperscript{3}University College London,\\ \textsuperscript{4}ENS, PSL University, \textsuperscript{5}Inria}

\date{}

\begin{document}

\newcounter{srCounter}
\newif\ifsrvar
\srvartrue
\ifsrvar
\newcommand{\seb}[1]{{\small \color{red} \refstepcounter{srCounter}\textsf{[SR]$_{\arabic{srCounter}}$:{#1}}}}
\else
\newcommand{\seb}[1]{}
\fi

\newcounter{fpCounter}
\newif\iffpvar
\fpvartrue
\iffpvar
\newcommand{\fabio}[1]{{\small \color{blue} \refstepcounter{fpCounter}\textsf{[FP]$_{\arabic{fpCounter}}$:{#1}}}}
\else
\newcommand{\fabio}[1]{}
\fi

\newcounter{apCounter}
\newif\ifapvar
\apvartrue
\ifapvar
\newcommand{\piktus}[1]{{\small \color{orange} \refstepcounter{apCounter}\textsf{[AP]$_{\arabic{apCounter}}$:{#1}}}}
\else
\newcommand{\piktus}[1]{}
\fi

\newcounter{plCounter}
\newif\ifplvar
\plvartrue
\ifplvar
\newcommand{\patrick}[1]{{\small \color{magenta} \refstepcounter{plCounter}\textsf{[PL]$_{\arabic{plCounter}}$:{#1}}}}
\else
\newcommand{\patrick}[1]{}
\fi

\newcounter{egCounter}
\newif\ifegvar
\egvartrue
\ifegvar
\newcommand{\egrave}[1]{{\small \color{purple} \refstepcounter{egCounter}\textsf{[EG]$_{\arabic{egCounter}}$:{#1}}}}
\else
\newcommand{\egrave}[1]{}
\fi

\newcounter{sbCounter}
\newif\ifegvar
\egvartrue
\ifegvar
\newcommand{\samuel}[1]{{\small \color{cyan} \refstepcounter{sbCounter}\textsf{[SB]$_{\arabic{sbCounter}}$:{#1}}}}
\else
\newcommand{\samuel}[1]{}
\fi

\newcounter{giCounter}
\newif\ifegvar
\egvartrue
\ifegvar
\newcommand{\gizacard}[1]{{\small \color{olive} \refstepcounter{giCounter}\textsf{[GI]$_{\arabic{giCounter}}$:{#1}}}}
\else
\newcommand{\gizacard}[1]{}
\fi

\newcommand{\citetodo}[1]{{\color{red}(#1)}}

\newcommand{\system}{\textsc{Side}}
\newcommand{\inferengine}{\emph{verification engine}}

\maketitle

\begin{abstract}
Verifiability is a core content policy of Wikipedia: claims that are likely to be challenged need to be backed by citations.
There are millions of articles available online and thousands of new articles are released each month. For this reason, finding relevant sources is a difficult task: many claims do not have any references that support them.
Furthermore, even existing citations might not support a given claim or become obsolete once the original source is updated or deleted.
Hence, maintaining and improving the quality of Wikipedia references is an important challenge and there is a pressing need for better tools to assist humans in this effort.
Here, we show that the process of improving references can be tackled with the help of artificial intelligence (AI).
We develop a neural network based system, called \system, to identify Wikipedia citations that are unlikely to support their claims, and subsequently recommend better ones from the web.
We train this model on existing Wikipedia references, therefore learning from the contributions and combined wisdom of thousands of Wikipedia editors.
Using crowd-sourcing, we observe that for the top 10\% most likely citations to be tagged as unverifiable by our system, humans prefer our system's suggested alternatives compared to the originally cited reference 70\% of the time.
To validate the applicability of our system, we built a demo\footnote{available at \url{https://verifier.sideeditor.com}} to engage with the English-speaking Wikipedia community and find that \system{}'s first citation recommendation collects over 60\% more preferences than existing Wikipedia citations for the same top 10\% most likely unverifiable claims according to \system{}.
Our results indicate that an AI-based system could be used, in tandem with humans, to improve the verifiability of Wikipedia.
More generally, we hope that our work can be used to assist fact checking efforts and increase the general trustworthiness of information online. All our code, data, indexes and models are publicly available at \url{https://github.com/facebookresearch/side}.
\end{abstract}

\section*{Introduction}

Wikipedia is one of the most visited websites on the web~\citep{similarweb}, and with half a trillion page views per year~\citep{wikistats}, constitutes one of the most important knowledge sources today. As such, it is critical that any knowledge on Wikipedia is \emph{verifiable}: Wikipedia users should be able to look up and confirm claims made on Wikipedia using reliable external sources~\citep{verifiability}. To facilitate this, articles provide inline citations that point to background material supporting the claim. Readers who challenge Wikipedia claims can follow these pointers and verify the information themselves~\citep{piccardi2020quantifying,lewoniewski2020modeling,kaffee2021references}. However, in practice this process can fail: a citation might either not entail the challenged claim, or its source might be questionable. 
Such claims may still be true, but a careful reader cannot easily verify them with the information at hand in the cited source. 
Under the assumption that a Wikipedia claim is true, its verification is hence a two stage process: 1) check the consistency of the existing source; 2) if that fails, search for new evidence, primarily online.

\begin{figure*}[t!]
    \centering
    \includegraphics[width=\linewidth]{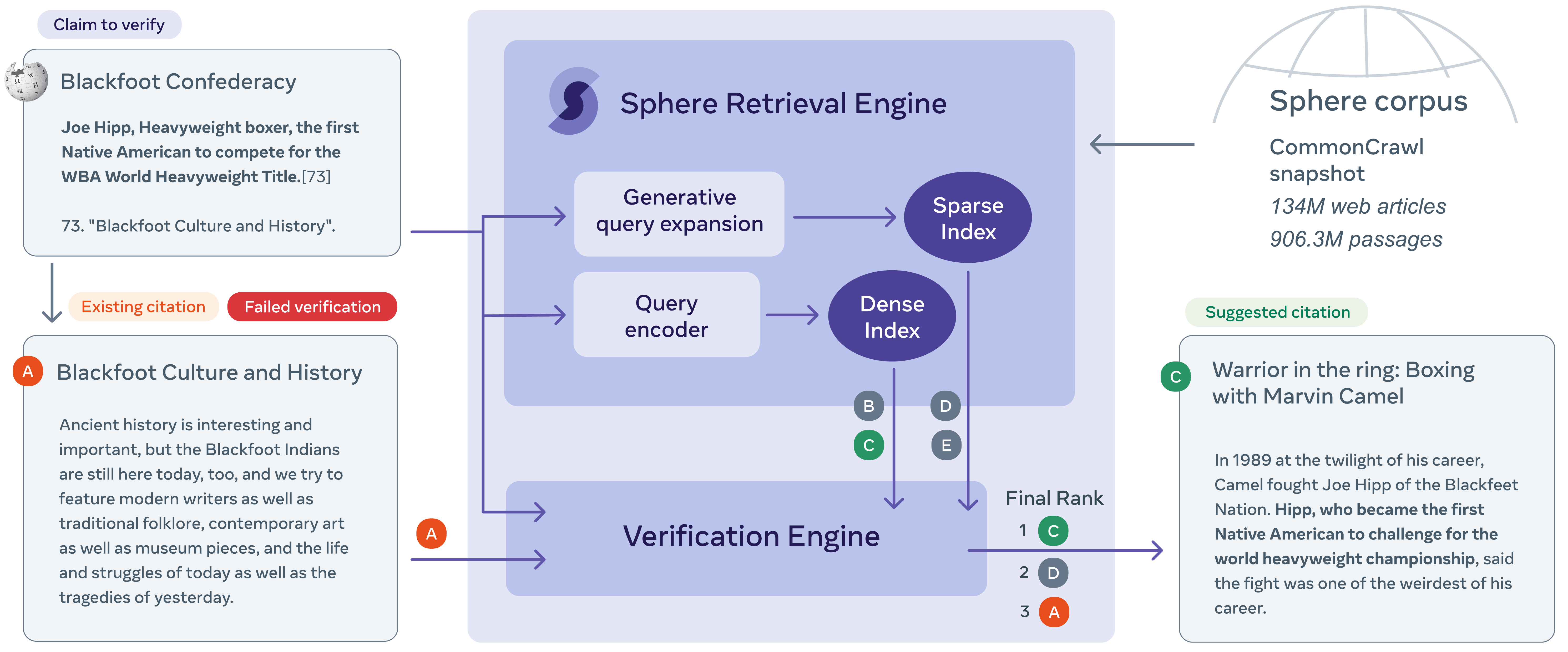}
    \caption{The decision flow of \system{} from a claim on Wikipedia to a suggestion for a new citation is as follows: (1) the claim is sent to the \emph{Sphere Retrieval Engine} which produces a list of potential candidate documents from the \textit{Sphere corpus}; (2) the \inferengine{} ranks the candidate documents and the original citation w.r.t. the claim; (3) if the original citation is not ranked above the candidate documents, then a new citation from the retrieved candidates is suggested. Note that the score of the \inferengine{} can be indicative of a potential \emph{failed verification}, as the one reported in the example.
}
    \label{fig:example}
\end{figure*}

Defined as above, verification of Wikipedia claims requires deep understanding of language and mastery of online search. To what extent can machines learn this behaviour? This question is important from the perspective of progress in fundamental AI. For example, verification requires the ability to detect logical entailment in natural language and to convert claims and their context to the best search term for finding evidence---two long-standing problems that have been primarily investigated in somewhat synthetic settings~\citep{bowman2015large,wang2018glue,camburu2018snli,nie2019adversarial,perez2017automatic,thorne2018fever,thorne2018automated}. It is equally important from a practical perspective. A machine verifier can assist Wikipedia editors by both flagging what citations might trigger failed verifications and suggesting what to replace citations with in case they currently do not support their respective claim. This can be significant: searching potential evidence and carefully reading the search results requires time and high cognitive effort. 
Integrating an AI assistant into this process could help to reduce both. 

In this work we develop \system{}, an AI-based Wikipedia citation verifier. \system{} finds claims on Wikipedia that likely cannot be verified given the current citation, and for such, scans a web snapshot for an alternative. Its behaviour is learnt using Wikipedia itself: using a carefully curated corpus of Wikipedia claims and their current citations, we train a) a retriever component that converts claims and contexts into symbolic and neural search queries optimised to find candidate citations in a web-scale corpus; and b) a verification model that ranks existing and retrieved citations according to how likely they might verify a given claim. 

We evaluate our model using both automatic metrics and human annotations. To measure the accuracy of our system automatically, we check how well \system{} recovers existing Wikipedia citations \emph{in high quality articles} as defined by the Wikipedia featured article class. We find that in nearly 50\% of the cases, \system{} returns exactly the source that is used in Wikipedia as its top solution. Notably, this does not mean the other 50\% are wrong but they are not what Wikipedia is currently using as a source. 

We also test \system's ability to be a citation assistant. In a user study we present existing Wikipedia citations next to the ones that \system{} produces. These users then assess to what extent the presented citations support the claim, and which citation---from \system{} or Wikipedia---would be better for verification. Overall, more than 60\% of the time users prefer \system{} citations over Wikipedia's ones, and this percentage grows above 80\% for cases in which \system{} associates a very low verification score to the Wikipedia citation. 

\section*{System Architecture}

In Figure~\ref{fig:example}, we provide a high level overview of \system{}  that shows an example of the decision flow given a Wikipedia claim. 
In the following, we briefly describe all major components of the system and how they interact with one another. Note that we use the term \emph{claim} to refer to the sentence (or clause) preceding a Wikipedia citation, but any given sentence can contain a multitude of logical claims, and the claim's meaning might depend on its context. 
The cited documents are represented as a list of passages, i.e., chunks of text with a fixed number of words.

\subsection*{The Retrieval Engine}
Given a claim tagged as \emph{failed verification} by a human editor, or flagged by our \inferengine, \system{} needs to retrieve a list of documents that support the claim. 
A human verifier would do so by 1) synthesizing a search query based on the claim's context; and 2) executing this query against a search engine. Fundamentally, \system{} \emph{learns} to do the same, using both sparse and dense retrieval sub-systems that we explain in more detail below. 
The claim's context is represented using the sentences preceding the citation, as well as the section title and the title of the enclosing Wikipedia article.
We use \textit{Sphere}~\citep{DBLP:journals/corr/abs-2112-09924}, a web-scale corpus and search infrastructure for web-scale data, as a source of candidate web pages.
Classic sparse and neural dense approaches are known to have complementary strengths~\citep{mao2020generation} and hence we merge their results to produce the final list of recommended evidence.           

The \emph{sparse retrieval} sub-system uses a seq2seq model~\citep{Lewis2019BARTDS,mao2020generation} to translate the citation context into query \emph{text}, and then matches the resulting query---a sparse bag-of-words vector---on a BM25 index~\citep{robertson1995okapi,baeza1999modern, schutze2008introduction, robertson2009probabilistic, lin2021pyserini} of Sphere. We train the seq2seq model using data from Wikipedia itself: the target queries are set to be web page titles of existing Wikipedia citations. In practice, we enrich the generated queries with the sentence preceding the citation and the Wikipedia title. The \emph{dense retrieval} sub-system is a neural network which learns from Wikipedia data to encode the citation context into a dense query vector~\citep{wu2019zero,karpukhin2020dense,maillard-etal-2021-multi, oguz2021domainmatched, 10.1162/tacl_a_00369}. This vector is then matched against the vector encodings of all passages in \textit{Sphere} and the closest ones are returned. The context and passage encoders are trained such that the context and passage vectors of existing Wikipedia citation and evidence pairs are maximally similar~\citep{karpukhin2020dense}.

\subsection*{The Verification Engine}
Given a claim and possible evidence document, either existing on Wikipedia or proposed by the retrieval engine, a human would carefully evaluate to what extent the claim is supported by the provided evidence. This is the role played by our \inferengine, a neural network taking the claim and a document as input, and predicting how well it supports the claim. Due to efficiency reasons, it operates on a per passage level and calculates the verification score of a document as the maximum over its per-passage scores. 
The verification scores are calculated by a fine-tuned BERT~\citep{devlin-etal-2019-bert} transformer that uses the concatenated claim and passage as input. This architecture is akin to prior work for textual entailment in natural language inference~\citep{maccartney-manning-2008-modeling}, i.e., testing whether a particular premise supports or contradicts a hypothesis. 

 The \inferengine{} is optimised to rank claim-document pairs in order of verifiability rather than making verified versus failed-verification decisions.  This is motivated by the way we envision \system{}'s usage in practical setting: we want to prioritise \emph{existing claims} for humans to check by starting with those that are \emph{less likely} supported by their current evidence, and to highlight \emph{recommended evidence} for a given claim by starting with documents that are \emph{more likely} to support the claim. To train the \inferengine{} model, we use a training objective that rewards models when they rank existing Wikipedia evidence higher than evidence returned by our retrieval engine. Assuming that some existing Wikipedia evidence is of poorer quality---a core motivation behind this work--- even though this training signal could be noisy, we found that, on average, it still provides a meaningful signal. We test this empirically further in the next section.

\section*{Evaluation and results}

Evaluating the performance of our system is challenging because we cannot be certain that existing citations are always accurate and because of the lack of annotations for citations that fail verification.
Therefore, we first evaluate the components of our system in isolation by addressing the following two questions: 1) given a Wikipedia claim, can our retrieval solutions surface the existing citation source from more than $100$M web articles? and 2) Is our \inferengine{} able to assign low scores to citations marked as failing verification in Wikipedia? After investigating these two questions, we conduct a large scale human annotation campaign to evaluate the overall system.

\subsection*{Experimental Data and Setting}

\begin{figure*}[ht]
    \centering
    \includegraphics[width=0.9\linewidth]{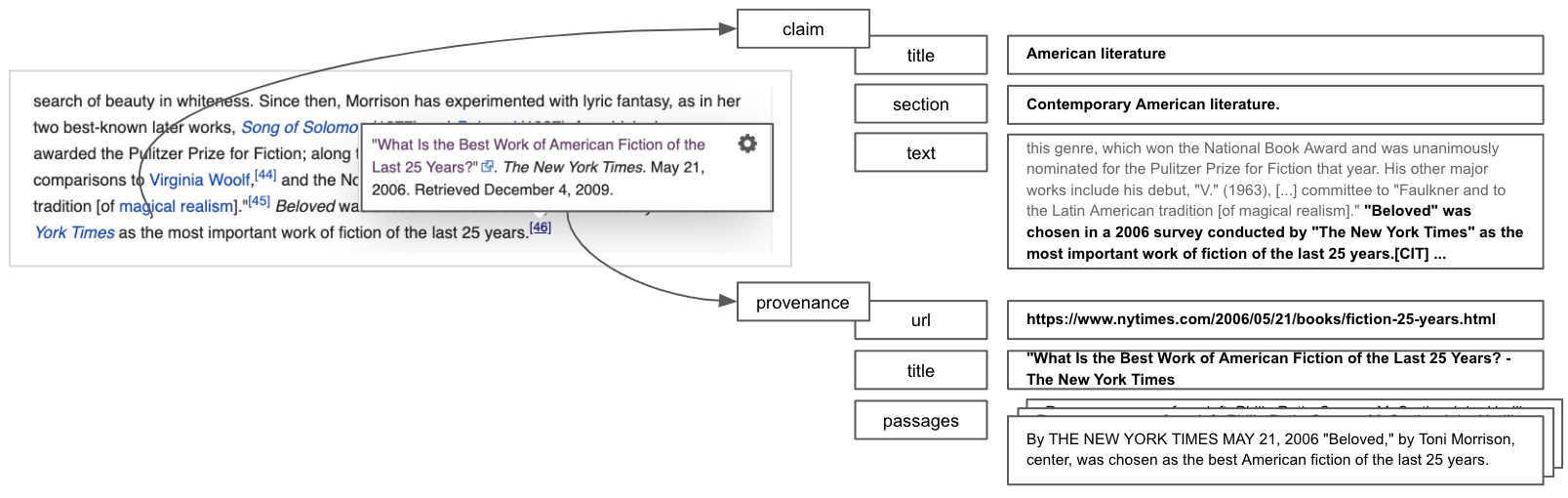}
    \caption{Example citation from the WAFER dataset.}
    \label{fig:example_citation}
\end{figure*}

We collect WAFER, a large scale dataset of English Wikipedia inline citations ($\approx3.8M$ instances - see \cref{tab:WAFERstats} for  statistics) which are aligned to a snapshot of the web to obtain the full textual content of the cited sources. 
Each instance in WAFER contains metadata from the claim's article, the text around the citation within the article (with a marker indicating the citation position), and metadata of the cited source, including title and full textual content (see Figure \ref{fig:example_citation} for an example). 
We create a cross-validation split on the article level---not on the citation level---to avoid potential test leakage into the training data.

Both the Wikipedia snapshot we consider (\ie, from KILT~\citep{petroni-etal-2021-kilt}) as well as the web snapshot (\ie, a CCNet~\citep{wenzek2019ccnet} dump from Sphere~\citep{DBLP:journals/corr/abs-2112-09924} which contains 134M web articles, split into 906.3M passages) are from August 2019. 
We use Sphere's web snapshot as the corpus for retrieval.
Aligning the citations in the Wikipedia snapshot and Sphere's web snapshot leads to $\approx250$k retrievable citations.
From those we sample $\approx4.5$k for testing and development each, making all the cited documents in our \textit{test} and \textit{dev} sets retrievable from the Sphere corpus. 
To increase the size of the training data, we match the remaining unaligned citations in the Wikipedia snapshot against several other Common Crawl snapshots from 2017 to 2019, collecting an additional $\approx3.5$M citations which are not retrievable from the Sphere corpus but which can be used for training models.

We distinguish two types of Wikipedia articles: \emph{featured articles}~\citep{featured} and \emph{non-featured articles}. 
Featured articles are a tiny fraction (\ie, $0.09$\%) of articles chosen by Wikipedia's editors as examples for their high quality.
Therefore, we use the featured articles only for evaluation given their limited number ($\approx 16$\% of test and dev citations). 
The remaining instances of the evaluation data are sampled from non-featured articles which can vary in quality in terms of writing or verifiability. We do not include in these datasets citations marked with a \textit{failed verification} template~\citep{failed}, which indicates that the source does not support what is claimed in the Wikipedia article.
We set these citations aside in specific dev and test sets (\ie, \textit{fail-dev} and \textit{fail-test}) in order to evaluate the ability of models to detect citations that fail verification.


We use popular retrieval metrics to measure the performance to rank the gold-cited document as high as possible in the retrieved results.
As our retrieval is passage-based, the highest ranked passage of a document determines its rank. 
We consider \textit{precision-at-1} (P@1), that is the percentage of evaluation instances in which the originally cited document was ranked in the first position among all retrieved documents.
Additionally, we use \textit{success-rate-at-k} (SR@k)---sometimes also referred to as HITS@k---which is the percentage of cases in which the originally cited document was amongst the top-k documents. 
We also use the Precision-Recall curve which measures the performance in terms of Precision when Recall is fixed to a certain level.

\begin{figure*}[t!]
\begin{subfigure}{.32\textwidth}
  \centering
  \includegraphics[width=\linewidth]{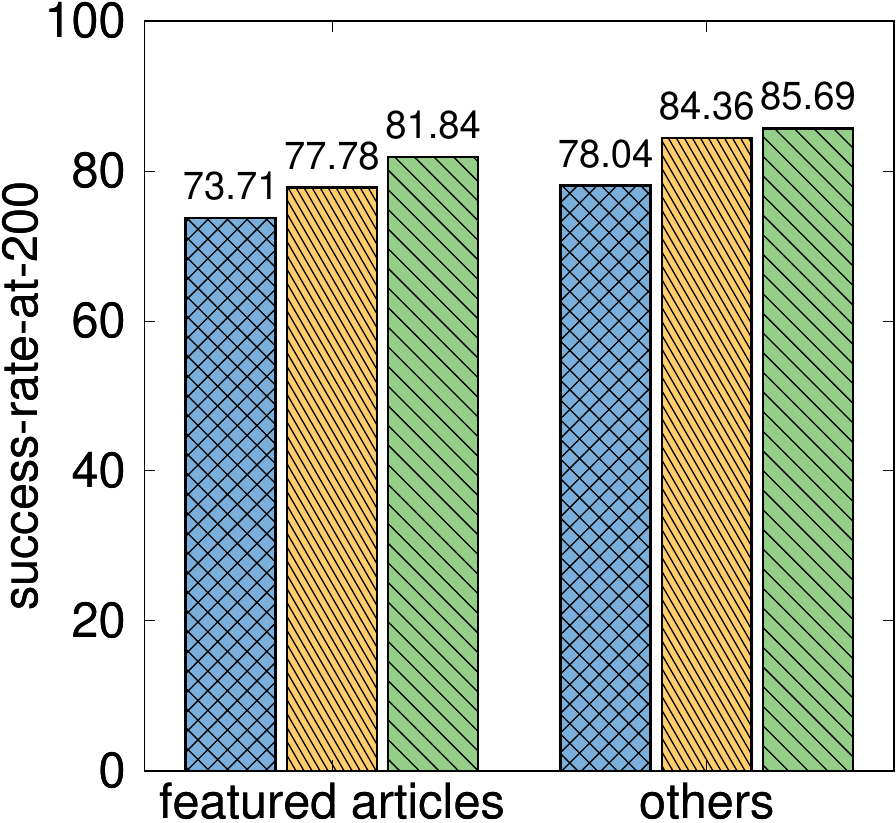}  
  \caption{Percentage of times our retrievers can surface the gold source among the top-200 results, for citations in featured and other Wikipedia articles. The \inferengine{} bar (\ie, green) combines sparse and dense retrievers, 100 passages each.}
  \label{fig:successrate}
\end{subfigure}
\hspace{.02\textwidth}
\begin{subfigure}{.32\textwidth}
  \centering
  \includegraphics[width=\linewidth]{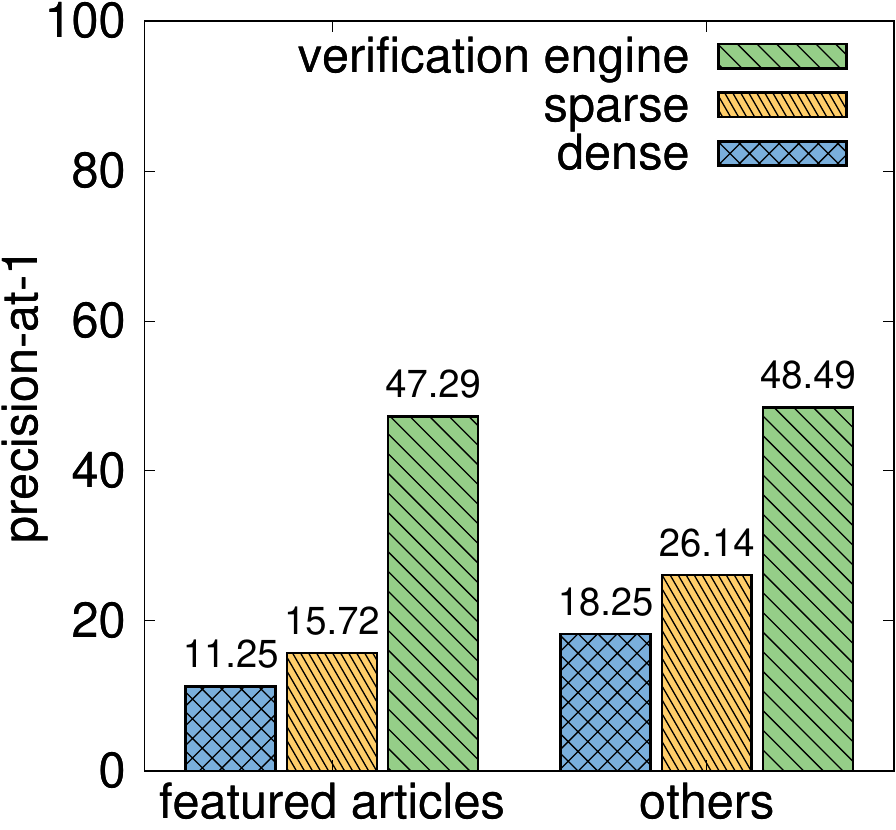}  
  \caption{Accuracy in surfacing the gold source in first position, for citations in featured and other articles. The \inferengine{} (\ie, green bar) takes as input a combination of 100 passages from the sparse and 100 from the dense retriever and reranks those.}
  \label{fig:precision}
\end{subfigure}
\hspace{.02\textwidth}
\begin{subfigure}{.31\textwidth}
  \centering
  \includegraphics[width=\linewidth]{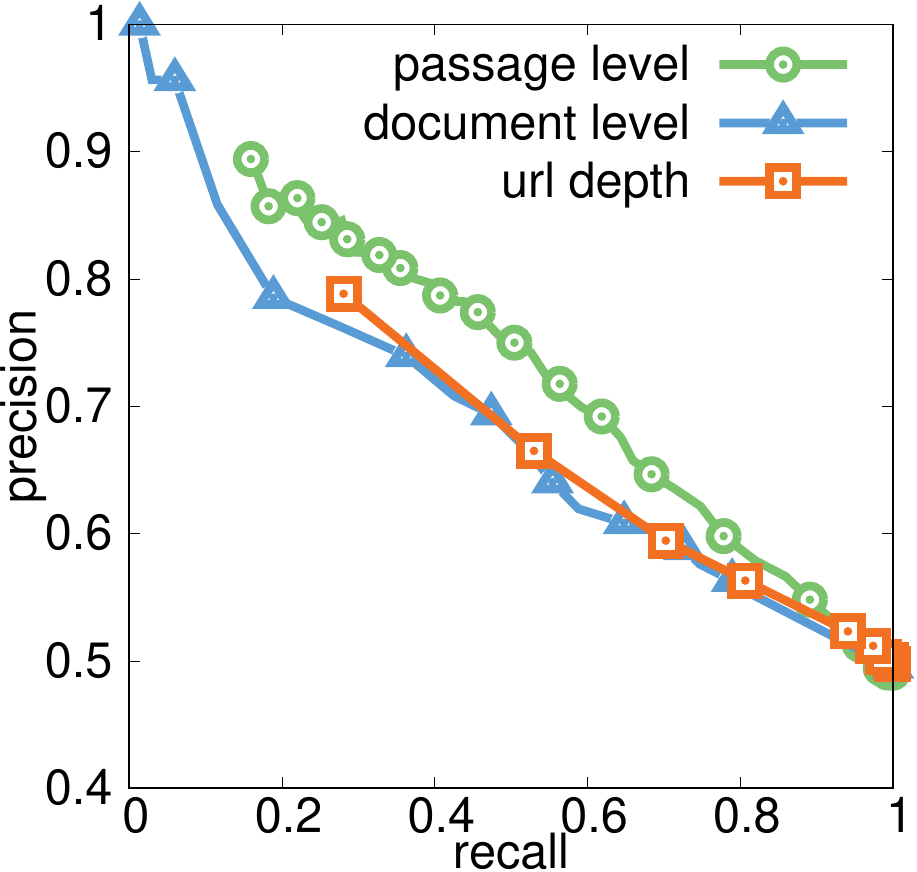}  
  \caption{Precision versus recall in detecting citations marked as \textit{failed verification} against citations in \textit{featured} articles. We compare a passage versus a document-level approach for the \inferengine{} and a baseline using the depth of the cited url. }
  \label{fig:RPcurve}
\end{subfigure}
\caption{Automatic evaluation of \system{} components on the WAFER test set.}
\label{fig:retrieval}
\end{figure*}

\subsection*{Retrieval evaluation}

We report our results in Figure \ref{fig:retrieval}. 
We note that the sparse retrieval solution outperforms the dense approach for retrieval from the web, which is consistent with previous observations~\citep{DBLP:journals/corr/abs-2112-09924}. 
However, 
we obtain our best overall SR@200 by combining 100 results from each given they are are highly complementary~\citep{mao2020generation} (see Figure \ref{fig:successrate}) --- this ensemble is what we use to retrieve passages to feed into the \inferengine{} component. 
Notably, the \inferengine{} component surfaces the original citation document  in the highest-ranked position nearly 50\% of the time (see Figure \ref{fig:precision}). 
However, these numbers have to be interpreted in the context of our background corpus, i.e., despite containing $\approx900$M passages from more than $100$M documents, it can only approximate a real-world scenario where evidences are to be sought on the open web. 

In general, retrieving evidence for claims in featured articles is more challenging than for other claims in Wikipedia, e.g., we observe a large difference of -7.0\%/-10.4\% P@1 (for dense/sparse) between featured and non-featured articles. 
One hypothesis for this is that there exists an intrinsic popularity bias associated with featured content. 
Featured content might often correlate with popular topics, which in turn means that more sources on the web contain relevant information.
In contrast, claims in more niche articles have much less coverage on the web and therefore are easier to find. 
Another factor is that featured articles are typically edited a lot more frequently, which is how they achieved their high quality, which in turn also could lead greater  deviation from the original phrasing of the cited source.
Assuming that dense retrievers are better at recognising paraphrases, we would expect a smaller increase in performance between featured and non-featured for dense vs. sparse, which is indeed the case.

The \inferengine{} model considerably boosts the accuracy of the retrieval component and almost levels the gap for featured articles, suggesting greater ability to identify evidence, even among a large set of relevant documents.
This performance can be explained by its ability to leverage fine-grained language comprehension, when the model can directly compare the contents of the two texts using a cross-attention mechanism to overcome the representational decomposability gap suffered by the retrievers~\citep{seo-etal-2019-real}.
Another relevant factor is that simple, helpful indicators like quoted phrases from the cited source seem to be easier to detect in  token-level comparison.

\begin{figure*}[t!]

\begin{subfigure}{.47\textwidth}
  \centering
  \includegraphics[width=\linewidth]{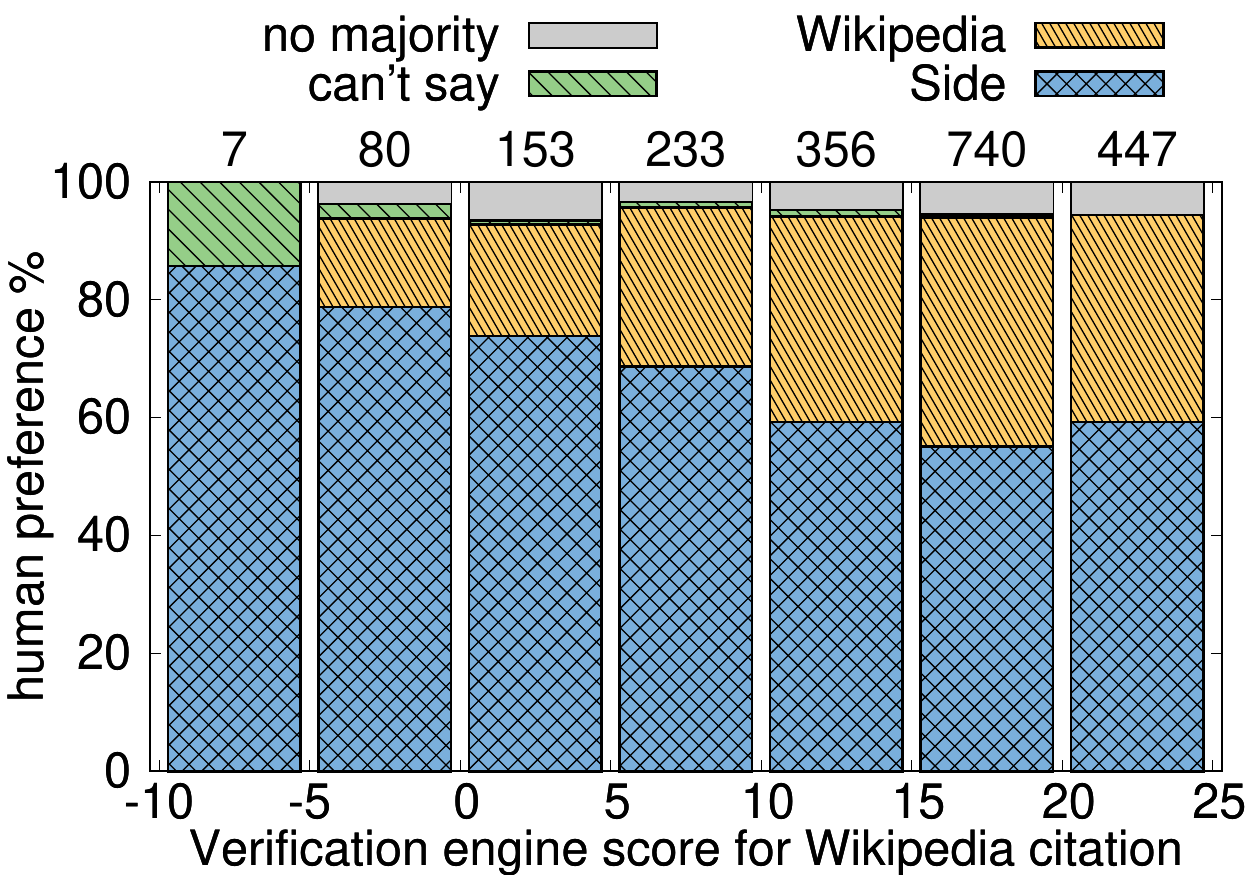}  
  \caption{Crowd annotators preference for citations suggested by \system{} versus those on Wikipedia for a given claim, without knowing their identity. Fleiss’ kappa Inter-Annotator Agreement = $0.2$.}
  \label{fig:pw}
\end{subfigure}\hspace{.02\textwidth}
\begin{subfigure}{.47\textwidth}
  \centering
  \includegraphics[width=\linewidth]{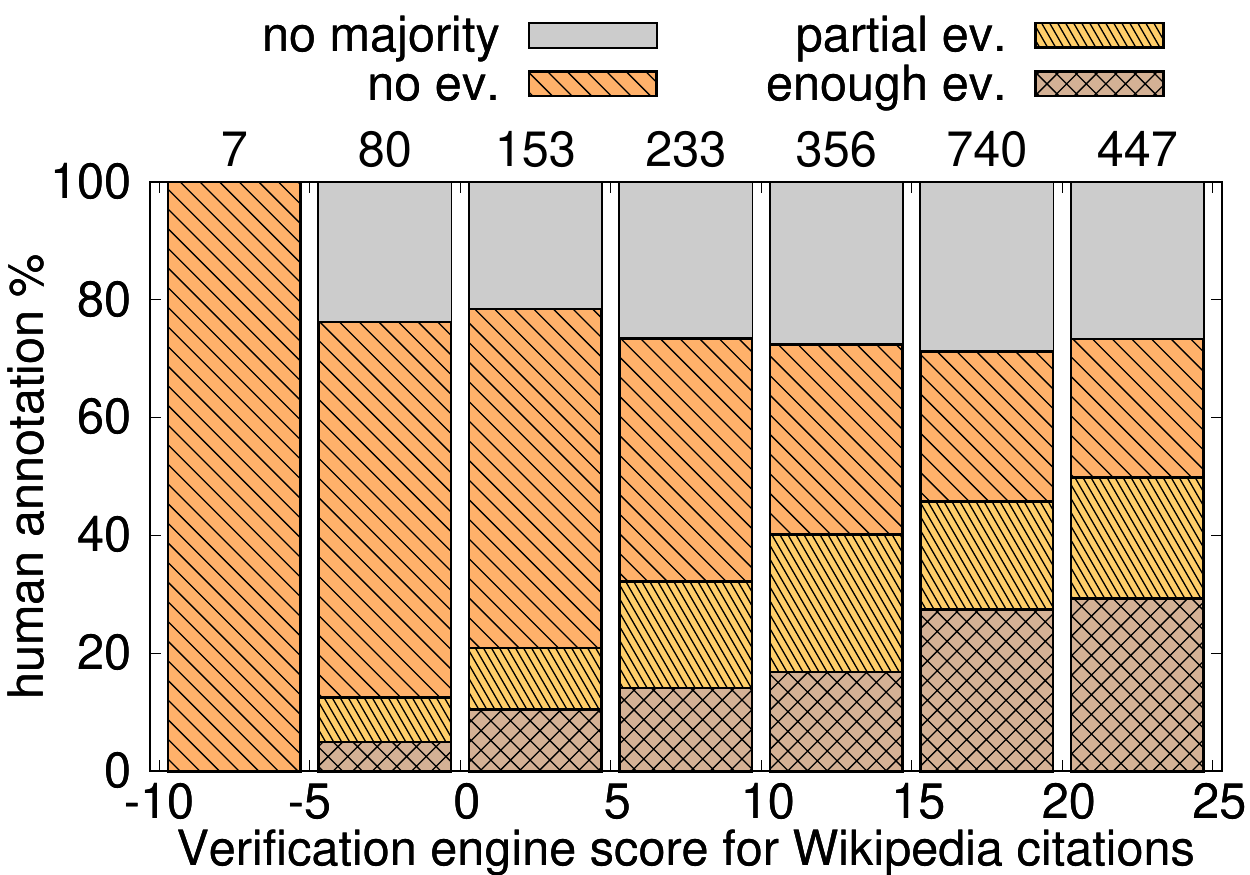}  
  \caption{Evidence annotations for Wikipedia citations: (1) \textit{enough} to verity the claim; (2) the claim is only \textit{partially} verified; (3) \textit{no evidence}. Fleiss’ kappa Inter-Annotator Agreement = $0.11$.}
  \label{fig:we}
\end{subfigure}
\caption{Crowd annotator evaluation for 2016 claims in the WAFER test set for which \system{} produces a citation with higher evidence score than the existing Wikipedia citation. We collect 5 annotations per claim and report majority voting results, bucketed according to the \inferengine{} score associated with the existing Wikipedia citation (bucket size reported on top).}
\label{fig:mturk}
\end{figure*}

\subsection*{Detecting Failed Verification}

Our goal in this analysis is to measure to which degree the score of the \inferengine{} can be used to detect whether a citation fails verification. To this aim, we rank the union of test citations in featured articles and test \textit{failed verification} citations. An ideal system would place all failed verification at the bottom end of the ranked list and featured citations at the top.
To compute the rank, we consider two different instantiations of the \inferengine, that operate either at a passage or document level. As many failed citations include a link to an over-generic URL (\eg, a generic newspaper website instead of a specific page covering the claim), we include a simple baseline based on the depth of a source URL (\ie, the number of elements in an URL path). In the passage-level solution, we independently compute a score for each passage in a document with the \inferengine{} and rank citations according to the maximum score. For the document-level approach, we feed as much text as possible (\ie, on average the first 2 or 3 passages) for the source document as input to a seq2seq model~\citep{Lewis2019BARTDS} and use the prediction score for the ranking. 

\begin{figure*}[t!]
    \centering
    \includegraphics[width=0.85\textwidth]{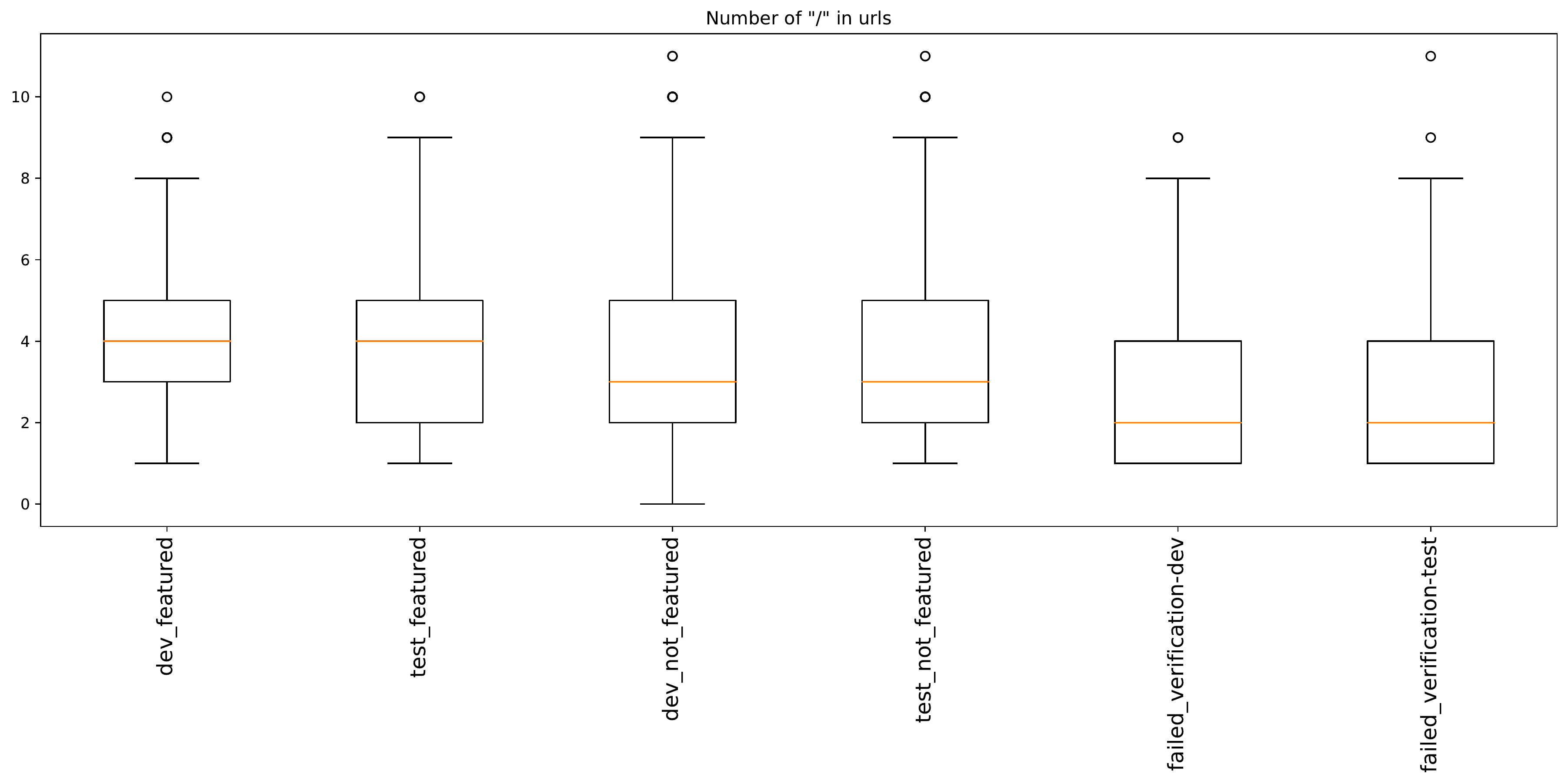}
    \caption{Url depth analysis.}
    \label{fig:depth}
\end{figure*}

The resulting precision-recall curve is in Figure \ref{fig:RPcurve}.
Overall, the passage-level \inferengine{} performs very well; if we only consider a conservative Recall of $15$\%, for instance, $\approx 90$\% are \textit{failed verification} citations. 
Notably, these results are achieved without any explicit supervision on failed verification instances, given that the \inferengine{} is trained only on positive examples. 
A document-level approach leads to worse results (\ie, $\approx 80$\% precision at $15$\% recall), mainly due to the impossibility of considering the whole document (given model architectural constraints on maximum input size). Considering url depth turns out to be a remarkably solid baseline. To further investigate this aspect, we study the distribution of depths for urls in our data (see Figure \ref{fig:depth}) and find that citations in featured articles tend to be deep (\ie, very specific urls) while citations marked as failed verification are usually shallow (\ie, very generic urls).

\subsection*{Evaluation of the final system}
To test the performance of our final system, we perform a two-stage human assessment: (1) a large scale crowd annotation campaign followed by (2) a smaller scale fine-grained evaluation. 
First, we select claims in the \textit{test set} for which \system{} outputs a citation source with a higher score than what is currently on Wikipedia. We then ask crowd annotators to express their preference on which one of the two (i.e., \system{}'s suggested citation or Wikipedia's one) better supports a given claim. Additionally, we ask them to assess if a source contains \textit{enough evidence} to support the claim, \textit{partial evidence} (meaning that only parts of the claim are supported by the source), or \textit{no evidence} whatsoever. To keep the annotation load tractable, we use our \inferengine{} component to select a single passage from each source, making sure to consider overlapping passages for Wikipedia sources so as to avoid cutting evidentiary sentences.

Results are reported in Figure \ref{fig:mturk}. We note that both preferences for \system{}'s suggested source (\ie, Figure~\ref{fig:pw}) and Wikipedia evidence annotations (\ie, Figure~\ref{fig:we}) are proportional to the ranker score associated to the existing Wikipedia citation---the lower the score the more preferences for \system{} and the less evidence found within Wikipedia. These results suggest that the ranker score might be a valid proxy for the presence (or absence) of evidence in a citation, and might help in surfacing cases that require attention from Wikipedia editors. To verify the noise introduced by automatically selecting a single passage for each source, we conduct a control study on more than 500 sources where we ask annotators if they prefer the selected passage (\ie, the top scored) with respect to a random one within the source. We find that for over $80\%$ of the cases annotators prefer the selected passage, with an Inter-Annotator Agreement of $0.27$ (Fleiss’ $\kappa$). Finally, to validate crowd annotators accuracy, we annotate more than 100 cases where evidence was not found in the Wikipedia citations. We find (see Table \ref{tab:expert} for the complete picture) that sometimes the evidence is in the source but not within the crawled text (\eg, multimedia content); other times, it is spread across multiple passages (which the current system can't detect, but that we plan to tackle in future work). Overall, more than $40\%$ of the time no evidence can be found in the reference to verify a Wikipedia claim.

\definecolor{nonecolor}{RGB}{122,195,106}
\definecolor{sidecolor}{RGB}{90,155,212}
\definecolor{wikicolor}{RGB}{254,194,70}

\begin{table}[t!]
\begin{minipage}{0.45\textwidth}
\centering
\begin{tabular}{ll}
\hline
\toprule
No evidence & 41.3\% \\
Partial evidence & 18.2\% \\
Full evidence in one passage & 16.7\% \\
Full evidence in multiple passages & 13.5\% \\
Evidence not in crawled text (e.g., multimedia) & 7.1\% \\
Pay wall access & 3.2\% \\
\bottomrule
\end{tabular}
\caption{Fine-grained human annotations for Wikipedia citations for which crowd annotators indicate no evidence for a total of 136 instances. }
\label{tab:expert}
\end{minipage}\hfill
\begin{minipage}{0.43\textwidth}
	\centering
	\begin{tikzpicture}[scale=0.48]
    \pie[
    color = {
        wikicolor!60, 
        sidecolor!60, 
        nonecolor!60, 
        gray!40},
    text = legend,
    explode = 0.1
    ]{16/Wikipedia,
    26/\system{},
    39/none of the two,
    19/no majority}
    \end{tikzpicture}
    
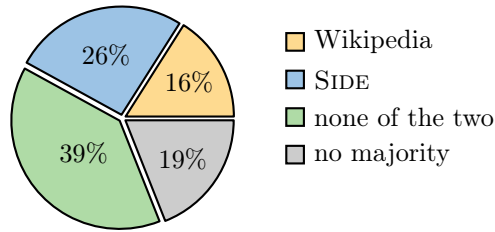
\captionof{figure}{Annotations  of Wikipedia authenticated users via our demo. P value = $0.0178$.  \label{fig:demo}}
\end{minipage}
\end{table}

Finally, we build a demo of \system{} and engage with the English-speaking Wikipedia community, asking users if they would use the citation already present on Wikipedia, the top-1 citation suggested by \system{} or none of the two to verify a given claim. We do not reveal the source of a citation in the user interface (i.e., Wikipedia or Side), select claim-citation pairs on Wikipedia that are likely to fail verification (i.e., with a verifier score below 0) and allow access to the full text for each citation (instead of a single passage).
Results (see Figure \ref{fig:demo}) reveal that \system{} can indeed select claim-citation pairs that fail verification --- users selected the Wikipedia citation in only $16\%$ of cases, compared to the $65\%$ of citations where either \system's recommendation or neither of the two were preferred. 
Moreover $26\%$ of the times \system{} can provide a top-1 recommendation that is judged appropriate by the community. We additionally conduct a sign test between \system{} and Wikipedia preferences resulting in a P value of $0.0178$ and two-tail P value of $0.0357$.
So far $43$ authenticated Wikipedia users\footnote{We exclude annotations performed by the authors of this paper.} participated to our study, for a total of $106$ annotations, with an average of $1.8$ annotations per claim. We plan to keep collecting annotations through our demo and update these figures in future iterations of the paper.
\section*{Related Work}

There is a large, passionate and engaged community who actively cares about, studies, and works to improve the verifiability of information in Wikipedia. 
The WikiProject Reliability~\citep{wikireliability}, for instance, contains a set of tools, resources and reports which are aimed at improving the reliability of Wikipedia articles.
One of these tools is Citation Hunt~\citep{citationhunt}, which  allows humans to check Wikipedia claims which have been flagged as not being backed by a reliable source and to propose a better citation. We believe the technology presented in this paper can be integrated with similar tools to surface more unverified claims and suggest potential alternative citations to a human to validate. 

Text-based classifiers able to detect claim needing citations~\citep{redi2019citation, chou2020citation} have received a lot of attention from both the scientific and the Wikimedia communities. We believe \system{} can be combined with such tools and recommend to Wikipedia editors a set of potential sources for claims needing a citation.
Several studies have also been conducted on user interactions with citations~\citep{piccardi2020quantifying,lewoniewski2020modeling,piccardi2021value,kaffee2021references,zagorova2022updated} that are tangential to our work.
There are a number of papers that approach citation recommendation for Wikipedia from different angles, such as by recommending citations from linked articles~ \citep{jana-etal-2018-wikiref} to citation span detection~\citep{fetahu-etal-2017-fine} amongst other efforts.
More broadly, citation retrieval and paper/source recommendation have also received attention in the scientific literature domain for many decades~\citep{mcnee,ren,bhagavatula-etal-2018-content,chou2020citation}, albeit with less of a focus on verifiability of existing citations, with citations drawn from much smaller and less diverse sources than the open web, see \citet{Frber2020} for a recent comprehensive review.

Several works have investigated the ability of AI to generate missing Wikipedia articles from scratch~\citep{liu2018generating,prabhumoye2019towards,fan2022generating,kaffee2022using}. There exists AI tools, such as~\cite{scribe}, that helps editors to bootstrap Wikipedia articles for underrepresented languages using these technologies. The \system{} engine can complement these systems and provide suggestions of supporting evidence from the web to back the article generation.

Finally, there exist a large body of research focused on fact-checking Wikipedia claims~\citep{thorne2018automated,thorne2018fever,thorne2019fever2,schuster-etal-2021-get,trokhymovych2022wikifactfind}. However, most of available resources are synthetically created to evaluate AI systems in a controlled environment. We believe that using real world supervision (\eg, from Wikipedia citations) could be key to unlock a larger applicability of these systems. 
\section*{Discussion}

We introduce \system, an AI-based system for improving the quality and verifiability of Wikipedia citations.
Building on recent advances in natural language processing, we demonstrate that machines can help humans finding better citations, a task requiring understanding of language, and mastery of online search.
While previous works~\citep{bowman2015large,wang2018glue,camburu2018snli,nie2019adversarial,perez2017automatic,thorne2018fever,thorne2018automated} have shown the ability of large neural networks to perform well on natural language understanding tasks, these results were mostly obtained for well specified tasks, on synthetic datasets specifically created for evaluating AI systems.
Here we show similar results in a real world scenario, implying noisier data and a more loosely defined task.

While our results are promising, and we believe our system could already be used to improve Wikipedia, there exist a variety of future research directions that can be pursued.
For instance, we only considered references corresponding to web pages, but Wikipedia also cites books, scientific articles and other kind of documents.
These include other modalities than just text, such as images and videos. To fully assess the quality of Wikipedia references, \system{} needs to become multi-modal.
Second, our system currently only supports the English language, while Wikipedia exists for more than two hundreds languages.
Making \system{} multi-lingual raises interesting research questions, such as the capabilities of performing \emph{cross-lingual} citation improvements:
given a claim in one language, if the system cannot find good evidence in that particular language, can it find references in other languages?

Finally, our work currently assumes that Wikipedia claims are verifiable, and only improves the quality of the references for existing claims.
A natural extension of our work would be to detect claims that are not verifiable, and flag them for review by human editors.
This comes with challenges, as a way to show that a claim is unverifiable is to find a contradicting evidence.
Unfortunately, Wikipedia currently does not contain such information, and thus training AI-based systems to perform this task is not straightforward.
However, we believe that \system{} could be a first step towards surfacing unverifiable claims: if \system{} cannot find good evidence for a claim, it might be impossible to verify.
We report one example of such claims in the Appendix (Table \ref{tab:example_no_evidence}), showing that a lack of good evidence from \system{} could be an indication of unverifiability.

We release all data, code and models described in this paper. 
We hope that this work could be used in a broader context than just Wikipedia, for example helping humans to perform fact-checking.
More generally, we believe that this work could lead to more trustworthy information online.
\section*{Acknowledgement}
The authors would like to greatly thank Miriam Redi and Diego Saez-Trumper from the Wikimedia Foundation for their invaluable help and support throughout the project; Ross Nkama, for helping set up our annotation interface; and all Wikipedia users who helped us evaluate \system{}.

\bibliography{sample}
\bibliographystyle{iclr2022_conference}

\section*{Additional information}

\begin{table*}[ht]
\centering
\resizebox{\linewidth}{!}{    
    \fontsize{8.4}{10.1}\selectfont
\begin{tabular}{r p{.7\textwidth}} \toprule
\multicolumn{2}{c}{Wikipedia content}
\\
\midrule
\textbf{Article} & \url{https://en.wikipedia.org/w/index.php?title=Jayda\%20Fransen&oldid=907222168} \\
\textbf{Input} & Jayda Fransen [SEP] Section::::Political career.:Leadership of Britain First. [SEP] she has often marched while holding a white cross, in "Christian patrols" through predominantly Muslim districts of British towns.
 In March 2018, Fransen was sentenced to 36 weeks imprisonment after being convicted of three counts of religiously aggravated harassment.
 Fransen had formerly been involved with the English Defence League, but left due to its association with violence.
 She was an unsuccessful candidate in the 2014 Rochester and Strood by-election, and the 2016 London Assembly election.
 Section::::Political career.
 Section::::Political career.:Leadership of Britain First.
Britain First, formed in 2011, is a British fascist political party founded by Paul Golding and Jim Dowson. Golding became the leader following the resignation of Dowson, and during this time Fransen was the deputy leader of the party. \textbf{Golding handed over the leadership role to Fransen in November 2016 due to his being sentenced to 2 months in prison for breaching a court order, although Fransen stated that his leave was in order "to address some important, personal family issues".[CIT]} Fransen stepped down from her leadership role in January 2019.
Section::::Political career.:Rochester and Strood by-election, 2014.
 Fransen stood as Britain First's first parliamentary candidate for the Rochester and Strood by-election on 20 November 2014, during which she expressed sympathy for the UK Independence Party (UKIP) and its candidate Mark Reckless (a Conservative MP who had switched allegiances to UKIP), who went on to win the seat.
 Britain First's campaign for the by-election drew attention when the party uploaded a photo of Fransen together with local activists from UKIP, who responded by saying  \\
 \midrule
   \multicolumn{2}{c}{Wikipedia citation}  \\
  \midrule
  \textbf{Source} & \url{http://www.searchlightmagazine.com/2016/12/more-questions-than-answers-a-searchlight-investigation} \\
  \textbf{Title} & More questions than answers: a Searchlight investigation \\
  \textbf{Passage} & brought against Golding? It came as no shock that Golding suddenly stood down from the leadership of Britain First on the first day of Mair’s trial, in favour of his deputy Jayda Fransen. When will Golding face a charge of incitement? When will somebody with responsibility and authority respond to these questions? Jayda Fransen with her Britain First \\
  \textbf{Score} & 2.6 \\
   \midrule
  \multicolumn{2}{c}{\system{} citation}  \\
   \midrule
  \textbf{Source} & \url{https://www.theguardian.com/uk-news/2016/nov/03/deputy-leader-britain-first-guilty-over-verbal-abuse-muslim-woman-jayda-fransen-hijab} \\
  \textbf{Title} & Deputy leader of Britain First guilty over verbal abuse of Muslim woman \\
  \textbf{Passage} & Deputy leader of Britain First guilty over verbal abuse of Muslim woman | UK news | The Guardian Deputy leader of Britain First guilty over verbal abuse of Muslim woman Far-right group’s Jayda Fransen convicted of religiously aggravated harassment for shouting at woman wearing hijab Thu 3 Nov 2016 13.19 EDT Last modified on Tue 28 Nov 2017 07.03 EST Jayda Fransen arriving at Luton magistrates court. Photograph: David Mirzoeff/PA The deputy leader of far-right group Britain First has been found guilty of religiously aggravated harassment after hurling abuse at a Muslim woman wearing a hijab in front of her four young children. Jayda Fransen, 30, was fined nearly £2,000 at Luton and South Bedfordshire magistrates court for'  \\
  \textbf{Score} & 9.97 \\
  \bottomrule
\end{tabular}
}
\caption{In this example, both Wikipedia and \system{} citation get a relatively low score from the \inferengine{}, suggesting the latter was unable to find enough evidence to verify the claim.}
\label{tab:example_no_evidence}
\end{table*}

\section*{Sphere retrieval}

\paragraph{Sparse retriever with generative query expansion}
Sparse retrieval methods rank documents by weighted lexical overlap and represent queries and documents as high-dimensional \emph{sparse} vectors with dimensions corresponding to vocabulary terms. 
BM25 is by nature very successful in retrieving passages that require high lexical overlap, also for long tail names and words.
The disadvantage for BM25 in this setting is that we do not know where the claim in the text in front of the citation is located, it could be just a short span of text, or the claim could be fragmented over multiple sentences and require references to the context of the Wikipedia article.
Indeed, in a manual evaluation of a small sample (30 instances) we found that roughly $1/3$ of the sentences had some kind of co-reference which was crucial for understanding the claim.  
Empirically we found that only using the first sentence in front of the claim and also adding the Wikipedia article's title to the query did yield the best BM25 results.

\paragraph{Dense retriever}
DPR is a method that learns to embed queries and documents as low-dimensional \emph{dense} vectors.
The basic building block of DPR is a BERT-like neural encoder, that consumes a sequence of tokens and predicts one dense vector. 
DPR consists of two such neural encoders, one for the query and one for a document's passage.
DPR is then trained on a dataset with instances consisting of (query, correct document) tuples.  
The training objective is to maximize the inner product between the query vector and the passage vectors of a correct document, and to minimize the inner product for incorrect documents. 
In contrast to BM25, DPR can learn which parts of the text are likely the important elements.
Another advantage is that DPR is typically stronger in retrieving passages with rephrased versions of the claim.

\paragraph{Training}
Many components of our system, such as the dense retriever and the \inferengine, are based on neural networks requiring examples to be trained.
We propose to leverage the scale of Wikipedia, and its millions of existing citations, to build a training set for our models.
It should be noted that the obtained data is noisy, as existing citations might be failing verification, and determining if it could be used to train our system is an interesting research question.
Moreover, our system processes references at the passage level, while our training data corresponds to pairs of claims and documents.
Thus, we train the retriever and the \inferengine{} using an expectation-maximization algorithm, modeling the passsage containing the evidence as a latent variable. 
Finally, our data only contains \emph{positive} examples of claims and references.
A standard solution for training retrievers is to mine \emph{negative} examples, and we follow this approach here.
While this work well for training retrievers, it is unclear how well this supervision would work for training the \inferengine, and in particular, to determine if an \emph{existing} reference is failing verification for a particular claims.
Indeed, the problem of ranking a set of candidate documents for a particular claim is different from ranking existing pairs of documents and claims.

\section*{Evaluation details}

\subsection*{CommonCrawl snapshots considered}
\label{sec:cc} 2017-26, 2017-39, 2017-51, 2018-13, 2018-26, 2018-39, 2018-51, 2019-18, 2019-30, 2019-43, 2020-05, 2017-30, 2017-43, 2018-05, 2018-17 ,2018-30, 2018-43, 2019-09, 2019-22, 2019-35, 2019-47, 2020-10, 2017-34, 2017-47, 2018-09, 2018-22, 2018-34, 2018-47, 2019-13, 2019-26, 2019-39, 2019-51

\begin{table}[ht]
\centering
\begin{tabular}{llll}
\hline
\toprule
split & size & articles & featured\\
\midrule
train & 3805958 & - & 0\\
dev & 4545 & - & 16\% (727)\\
test & 4568 & - & 16\% (738)\\
\midrule
fail-dev & 725 & - & 0 \\
fail-test & 730 & - & 0\\
\bottomrule
\end{tabular}
\caption{\label{tab:WAFERstats} Statistics for the WAFER dataset.}
\end{table}
\begin{table}[ht]
\centering
\resizebox{\textwidth}{!}{
\begin{tabular}{llllllllll}

 \toprule
  &  \multicolumn{3}{c}{featured} & \multicolumn{3}{c}{random} &  \multicolumn{3}{c}{micro avg.} \\
 & P@1 & SR@100 & SR@200 & P@1 & SR@100 & SR@200 & P@1 & SR@100 & SR@200 \\
\midrule
\multicolumn{10}{c}{1st stage - Sphere Retrieval} \\
\midrule

1. dense - DPR multi-task pretrained & 5.33 & 33.82 & - & 6.81 & 30.70 & - & 6.57 & 31.22 & - \\
2. dense - DPR from scratch & 11.25 & 66.94 & 73.71 & 18.25 & 72.09 & 78.04 & 17.12 & 71.26 & 77.34 \\
3. sparse - BM25 no expansion & 15.58 & 68.44 & - & 24.57 & 74.18 & - & 23.1 & 73.24 & - \\
4. sparse - BM25 with expansion & 15.72 & 73.17& 77.78 & 26.14 & 80.16 & 84.36 & 24.45 & 79.02 & 83.30 \\
\midrule
2. dense + 4. sparse & - & - & \textbf{81.84} & - & - & \textbf{85.69} & - & - & \textbf{85.07} \\
\midrule 
\multicolumn{10}{c}{2nd stage - Evidence Ranking} \\
\midrule 
\inferengine \  (2. dense + 4. sparse) & \textbf{47.29} & \textbf{81.71} & - & \textbf{48.49} & \textbf{85.46} & - & \textbf{48.29} & \textbf{84.85} & - \\
\bottomrule
\end{tabular}
}
\caption{\label{tab:wafertest} WAFER test results.}
\end{table}

\begin{figure*}[ht]

\begin{subfigure}{.47\textwidth}
  \centering
  \includegraphics[width=\linewidth]{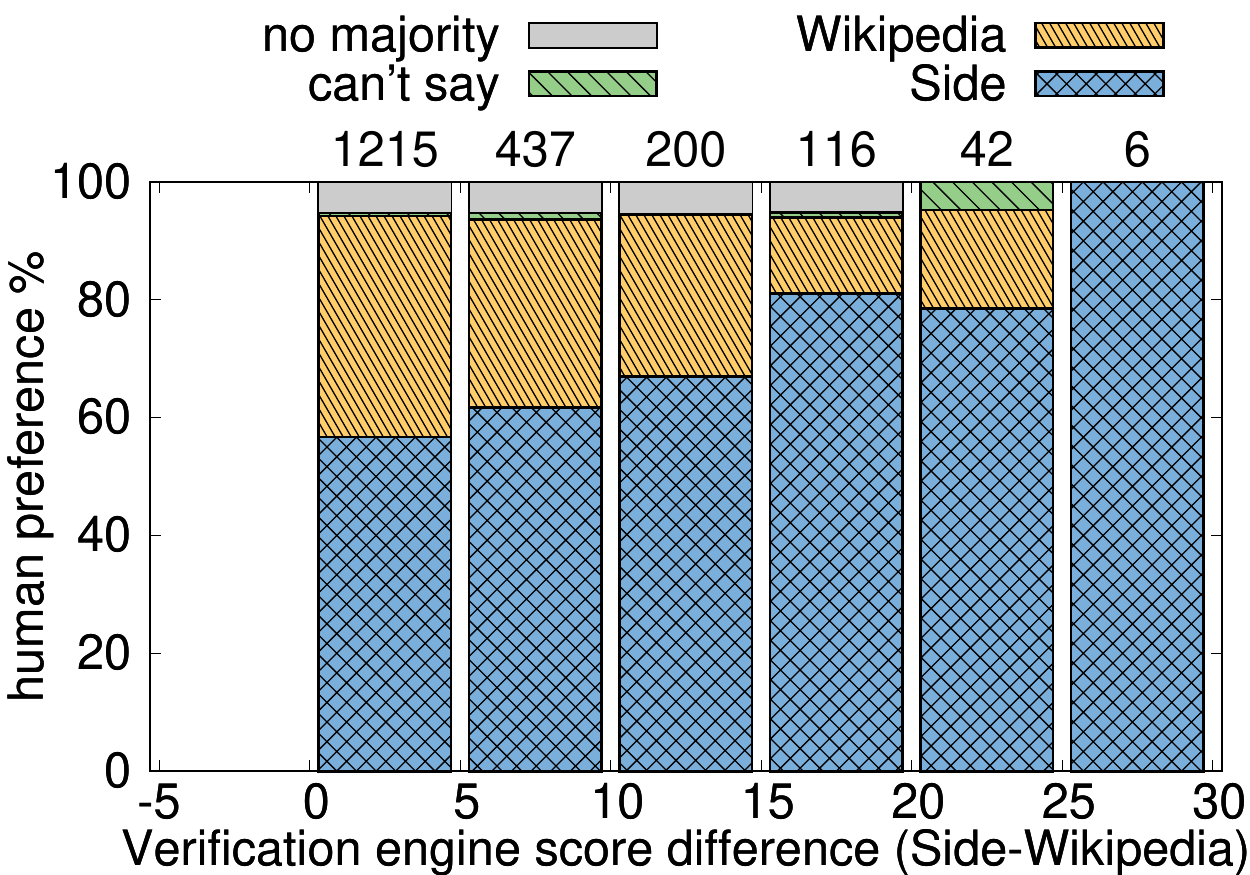}  
  \caption{Crowd annotators preference for citations suggested by our System versus those present in Wikipedia for a given claim. Fleiss’ kappa Inter-Annotator Agreement = $0.2$.}
  \label{fig:pd}
\end{subfigure}\hspace{.02\textwidth}
\begin{subfigure}{.47\textwidth}
  \centering
  \includegraphics[width=\linewidth]{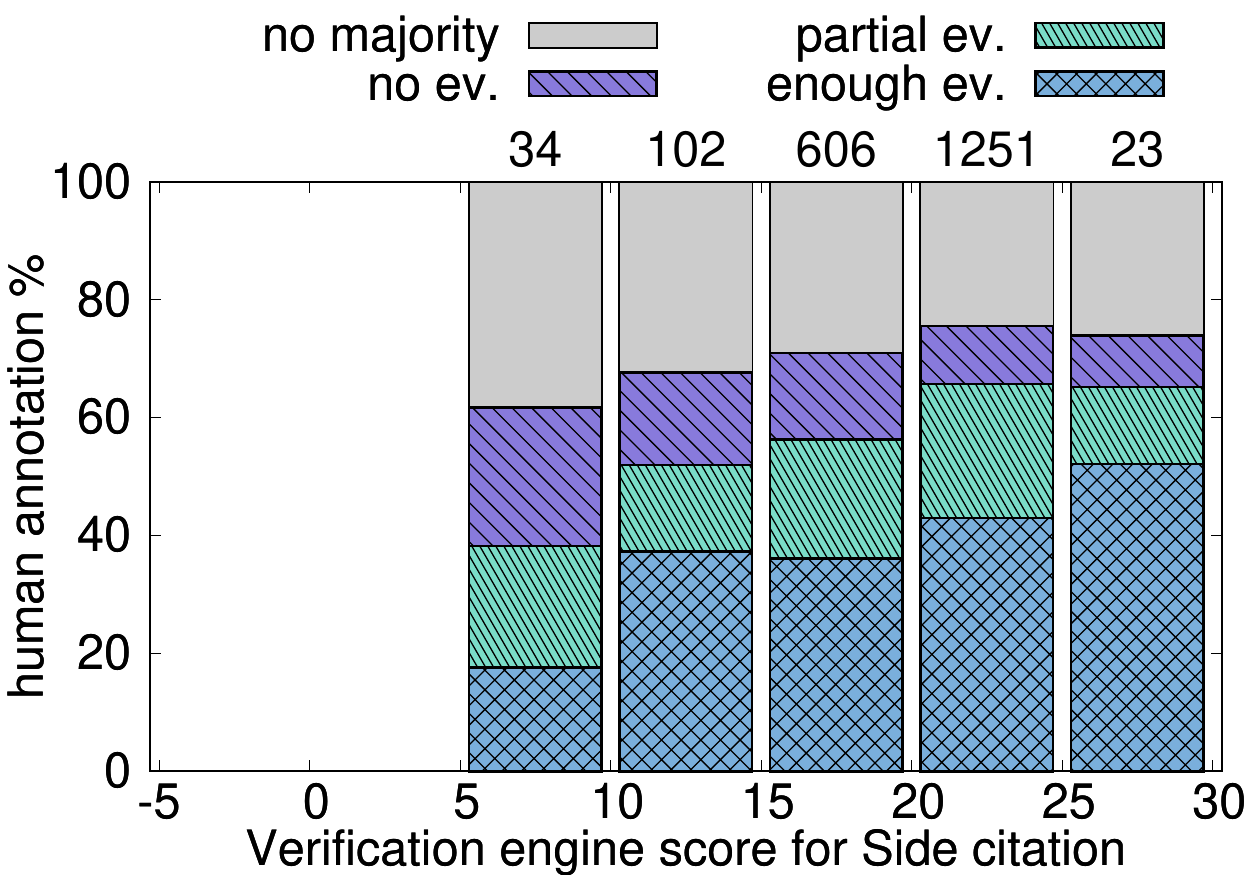}
  \caption{Evidence annotations for \system{} citations: (1) \textit{enough} to verity the claim; (2) the claim is only \textit{partially} verified; (3) \textit{no evidence}. Fleiss’ kappa Inter-Annotator Agreement = $0.09$.}
  \label{fig:se}
\end{subfigure}
\caption{Crowd annotators evaluation for 2016 claims in the WAFER test set for which \system{} produces a citation with higher evidence score that the existing Wikipedia citation. We collects 5 annotations per claim and report majority voting results, bucketed according to the evidence ranker score (bucket size reported on top).}
\label{fig:mturk2}
\end{figure*}

\end{document}